\documentclass[aps,pra,onecolumn,notitlepage,10pt,citeautoscript,longbibliography]{revtex4-1}
\usepackage{amssymb}
\usepackage{amsmath}
\usepackage[pdftex]{graphicx}
\usepackage{dcolumn}
\usepackage{bm}
\usepackage{textcomp}

\usepackage{xcolor}
\definecolor{trueblue}{rgb}{0.0, 0.45, 0.81}
\definecolor{crimsonglory}{rgb}{0.75, 0.0, 0.2}

\usepackage{hyperref}
\hypersetup{colorlinks,linkcolor={trueblue},citecolor={crimsonglory},urlcolor={crimsonglory}}

\begin{document}

\title{Magneto-optical conductivity of anisotropic two-dimensional Dirac-Weyl materials}

\author{M. Oliva-Leyva}
\email{moliva@iim.unam.mx} 
\author{Chumin Wang}
\email{chumin@unam.mx}

\affiliation{Instituto de Investigaciones en Materiales, Universidad Nacional Aut\'{o}noma de M\'{e}xico, Apartado Postal 70-360, 04510 Mexico City, Mexico.}


\begin{abstract}
In the presence of an external magnetic field, the optical response of two-dimensional materials, whose charge carriers behave as massless Dirac fermions with arbitrary anisotropic Fermi velocity, is investigated. Using Kubo formalism, we obtain the magneto-optical conductivity tensor for these materials, which allows to address the magneto-optical response of anisotropic Dirac fermions from the well known magneto-optical conductivity of isotropic Dirac fermions. As an application, we analyse the combined effects of strain-induced anisotropy and magnetic field on the transmittance, as well as on the Faraday rotation, of linearly polarized light after passing strained graphene. The reported analytical expressions can be a useful tool to predict the absorption and the Faraday angle of strained graphene under magnetic field. Finally, our study is extended to anisotropic two-dimensional materials with Dirac fermions of arbitrary pseudospin.


\end{abstract}


\maketitle

\section{Introduction}

A Dirac-Weyl material, such as graphene \cite{Novoselov2005,Zhang2005}, organic conductors \cite{Katayama2009,Kajita2014} and topological insulators \cite{Hasan2010,Qui2011}, possesses low-energy fermionic excitations that behave as massless Dirac particles, rather than conventional fermions governed by the Schr\"{o}dinger's equation \cite{Wehling2014}. The behaviour of these Dirac fermions in graphene has been studied by applying an external magnetic field, where a half-integer quantum Hall effect was observed \cite{Novoselov2005,Zhang2005}. This observation demonstrates the existence of relativistic Landau levels with a square root dependence on both the magnetic field $B$ and Landau level index $n$ (as $\sim\sqrt{B\vert n\vert}$), which is in stark contrast to the equally spaced Landau levels for a conventional two-dimensional electron gas. This unconventional Landau spectrum has also been proved by means of infrared spectroscopy measurements, whose transmittance through graphene under magnetic field is in excellent agreement with the theoretical magneto-optical response of Dirac fermions derived from the Kubo formula \cite{Sadowski2006,Jiang2007}. Moreover, graphene exhibits quantum Faraday and Kerr rotations associated with the half-integer quantum Hall effect \cite{Morimoto2009,Shimano2013}.

Even in absence of magnetic field the optical properties of graphene are \emph{per se} unusual. For example, graphene presents an universal transmittance $T$ determined by the fine-structure constant $\alpha$ (being $T\approx1-\pi\alpha\approx97.7\%$), over a broad range of frequencies \cite{Nair2008}. This remarkable feature is a consequence of its charge carriers behaved as massless Dirac fermions. At the same time, graphene exhibits a large interval of elastic response and then, mechanical deformations have been proposed as a tool to tune its optical properties \cite{Bae2013,Ni2014,Rakheja2016,Ngugen2017}. By applying a uniaxial strain, the optical conductivity of graphene becomes anisotropic \cite{Pellegrino2010,Pereira2010} and its transmittance depends on the incident light polarization \cite{Ni2014,Oliva2015}. 

Up to now, the combined effects of both magnetic field and strain on the optical properties of a two-dimensional Dirac-Weyl material (2D DWM) have not been analysed in detail. In fact, the optical conductivity of unstrained graphene under magnetic field is given by an antisymmetric tensor \cite{Gusynin2006}, whereas the optical response of strained graphene, in absence of magnetic field, is a symmetric tensor \cite{Oliva2014}. In consequence: \emph{What is the symmetry of the optical conductivity tensor if both effects are present? How many independent components does this tensor have?} 

The main objective of this article is to provide a general formulation of the magneto-optical conductivity for anisotropic (strained) 2D DWMs. For this purpose, in Section~\ref{SLL} we start by deriving the Landau level spectrum for the mentioned 2D DWMs. Unlike previous approaches \cite{Goerbig2008,Morinari2009}, our derivations are carried out in an arbitrary laboratory reference system. In Section~\ref{SecMOC}, we give an analytical expression for the magneto-optical conductivity tensor of an anisotropic 2D DWM,  while we answer the above questions in Section~\ref{Dis}. As an example, we apply our analytical results to a strained graphene and we report a generalized Faraday rotation. Section~\ref{Gen} is devoted to discuss the extension of this analysis to Dirac fermions with arbitrary pseudospin and, finally, some conclusions are given in Section~\ref{Con}.

\section{Landau levels}\label{SLL}

\begin{figure}[ht]
\centering
\includegraphics[width=0.9\linewidth]{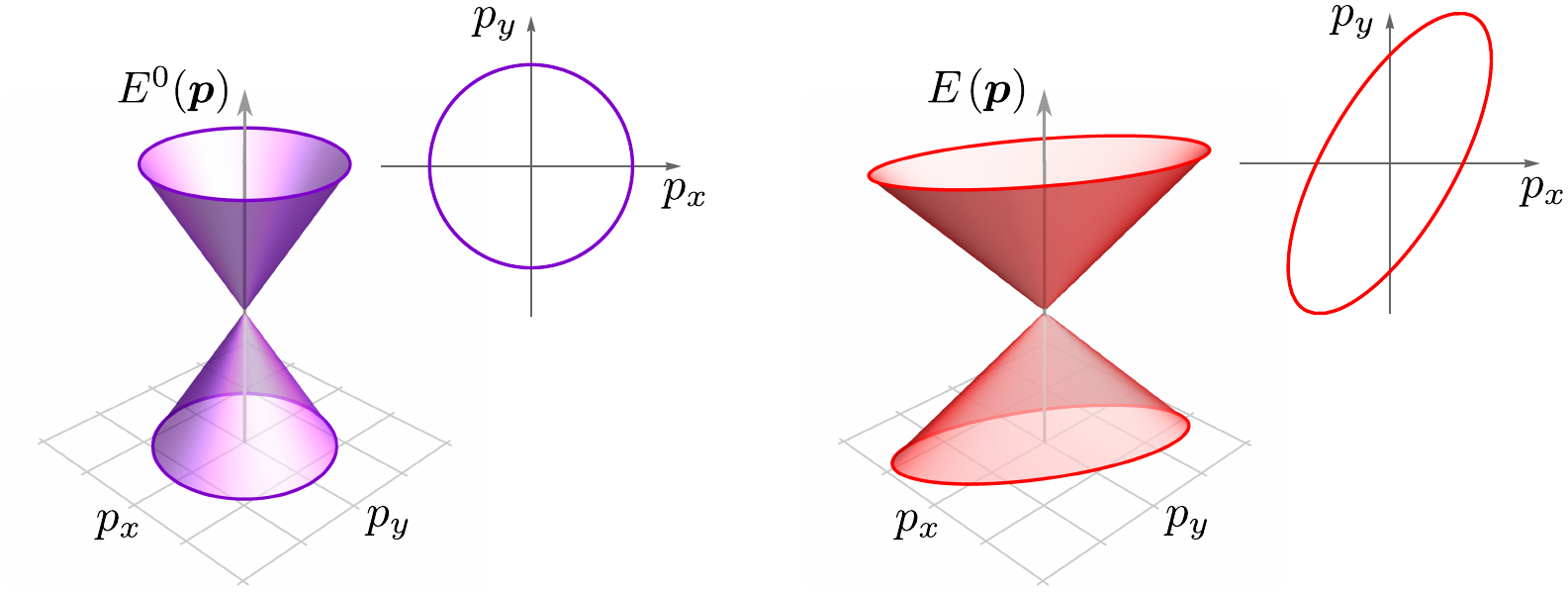}
\caption{\label{fig1} The left and right panels illustrate the energy dispersion relations $E(\bm{p})$ of an isotropic and an anisotropic 2D DWMs, respectively. The isoenergetic curves for the isotropic material are circles (violet contours) while for the anisotropic one are ellipses (red contours).}
\end{figure}

We consider the dynamics of low-energy carriers in an anisotropic 2D DWM governed by the generic Dirac-Weyl Hamiltonian \cite{Pellegrino2011,Oliva2013,Trescher2015,Hirata2016,Li2017}
\begin{eqnarray}\label{AH}
\mathcal{H}=\bm{\tau}\cdot\mbox{\bf{v}}\cdot\bm{p}=\sum_{i,j} \tau_{i}v_{ij}p_{j},
\end{eqnarray}
where $\bm{\tau}=(\tau_{x},\tau_{y})$ are the first two Pauli matrices that act on the pseudospin degree of freedom, $\bm{p}$ is the momentum measured from the Dirac point and $\mbox{\bf{v}}$ is the $(2\times2)$ symmetric Fermi velocity tensor. The corresponding energy dispersion relation is 
\begin{equation}\label{DR}
E(\bm{p})=\pm\sqrt{\sum_{i}\big(\sum_{j} v_{ij}p_{j}\big)^{2}},
\end{equation}
which represents elliptic Dirac cones. Unlike an isotropic 2D DWM, whose Hamiltonian is $\mathcal{H}^{0}=v_{0}\bm{\tau}\cdot\bm{p}$ with energy dispersion $E^{0}(\bm{p})=\pm v_{0}\vert\bm{p}\vert$, the constant energy contours of Eq.~(\ref{DR}) are not circles but ellipses, as illustrated in Fig.~\ref{fig1}. 

Since the Fermi velocity tensor is symmetric, such that $v_{ij}=v_{ji}$, one can choose a particular coordinate system $x'y'$ (the principal one) where $\mbox{\bf{v}}$ is diagonal and then, the Hamiltonian (\ref{AH}) reduces to $\mathcal{H}=\tau_{x'}v_{x'x'}p_{x'} + \tau_{y'}v_{y'y'}p_{y'}$, where $\tau_{x'}=\tau_{x}$ and $\tau_{y'}=\tau_{y}$ remain being the Pauli matrices. However, the anisotropic Dirac-Weyl Hamiltonian written in the form (\ref{AH}) allows us to obtain the general expression of the magneto-optical conductivity tensor an arbitrary laboratory reference system, which will lead to a more complete mathematical description of both anisotropy and magnetic field effects on the optical response. For example, the general expression of optical conductivity tensor for uniaxially strained graphene provides the possibility to determine the strain state by means of only two transmittance measurements using linearly polarized light with different polarization angles \cite{Oliva2015}. 

Now, in the presence of a uniform magnetic field $\bm{B}=B\bm{e}_{z}$ perpendicular to the 2D DWM sample, Hamiltonian (\ref{AH}) becomes according to the Peierls' substitution,
\begin{equation}\label{AHB}
\mathcal{H}=\bm{\tau}\cdot\mbox{\bf{v}}\cdot\bm{\Pi}=\sum_{i,j} \tau_{i}v_{ij}\Pi_{j},
\end{equation} 
where $\bm{\Pi}=\bm{p}+e\bm{A}$ is the gauge-invariant kinetic momentum, being $-e < 0$ the electron charge and $\bm{A}$ the vector potential associated to the magnetic field through $\bm{B}=\bm{\nabla}\times\bm{A}$. The components of $\bm{\Pi}$ satisfy the commutation relation $[\Pi_{x},\Pi_{y}]=-i e \hbar B$ \cite{Goerbig2011}. In general, we can introduce the following ladder operators,  
\begin{equation}
a= \frac{v_{xx}\Pi_{x}+v_{xy}\Pi_{y} - i(v_{yy}\Pi_{y}+v_{xy}\Pi_{x})}{\sqrt{2e\hbar B \mbox{det}(\mbox{\bf{v}})}}
\end{equation}
and
\begin{equation}
a^{\dagger}= \frac{v_{xx}\Pi_{x}+v_{xy}\Pi_{y} + i(v_{yy}\Pi_{y}+v_{xy}\Pi_{x})}{\sqrt{2e\hbar B \mbox{det}(\mbox{\bf{v}})}},
\end{equation}
such that $[a,a^{\dagger}]=1$, in analogy to those defined in Refs. \cite{Goerbig2008,Morinari2009}. In terms of these ladder operators, the Hamiltonian (\ref{AHB}) can be rewritten as 
\begin{equation}\label{}
\mathcal{H}=\sqrt{2e\hbar B\mbox{det}(\mbox{\bf{v}})}
\left(\begin{array}{cc}
0 & a\\
a^{\dagger} & 0
\end{array}
\right),
\end{equation}
whose eigenvalues obtained from $\mathcal{H}\vert\psi_{n}\rangle=E_{n}\vert\psi_{n}\rangle$ are
\begin{equation}\label{LL}
E_{n}=\mbox{sgn}(n)\sqrt{2e\hbar B\mbox{det}(\mbox{\bf{v}})\vert n\vert}\ \ \ \mbox{with}\ n\in\mathbb{Z}. 
\end{equation}
The corresponding eigenspinors are given by 
\begin{equation}
\vert\psi_{n}\rangle = \frac{1}{\sqrt{2}}
\left(\begin{array}{c}
\vert\hspace{0.5mm}\vert n\vert -1\hspace{0.5mm} \rangle\\
\mbox{sgn}(n) \vert\hspace{0.5mm}\vert n \vert\hspace{0.5mm}\rangle
\end{array}\right)\ \ \ \mbox{for}\ n\neq0, 
\end{equation}
while the zero-energy eigenspinor is
\begin{equation}
\vert\psi_{0}\rangle=
\left(\begin{array}{c}
0\\
\vert 0 \rangle
\end{array}\right), 
\end{equation} 
where $\vert\hspace{0.5mm}\vert n\vert\hspace{0.5mm}\rangle$ are the eigenstates of the usual number operator $a^{\dagger}a$ \cite{Goerbig2011}.  

Let us draw attention to two important remarks about the Landau level spectrum (\ref{LL}). First, the energies $E_{n}$ are independent of any choice of the coordinate system, as physically expected, because $E_{n}$ are expressed as a function of $\mbox{det}(\mbox{\bf{v}})$, which together with $\mbox{tr}(\mbox{\bf{v}})$ are the two principal invariants of the $(2\times 2)$ tensor $\mbox{\bf{v}}$ under rotations. Note that using the identity $\mbox{det}(\mbox{\bf{v}})= v_{x'x'}v_{y'y'}$, Eq. (\ref{LL}) reproduces the Landau levels reported for anisotropic 2D DWM in previous works \cite{Goerbig2008,Morinari2009}. Second, an isotropic 2D DWM with Hamiltonian $\mathcal{H}^{0}=v_{0}\bm{\tau}\cdot\bm{p}$, hereafter used as reference material, has exactly the same Landau level spectrum (\ref{LL}) if, instead of being under a magnetic field $B$, it is in an effective magnetic field of magnitude
\begin{equation}\label{EB}
\mathfrak{B}=B\mbox{det}(\mbox{\bf{v}})/v_{0}^{2}.
\end{equation} 
This fact will be useful for calculating the magneto-optical conductivity of anisotropic 2D DWMs.

\section{Magneto-optical conductivity}\label{SecMOC}

The optical conductivity tensor $\bm{\sigma}(\omega,B)$ of a 2D electron system in a magnetic field can be studied using the Kubo formalism \cite{Mahan}. In the Landau-level representation, the Kubo formula reads \cite{MacDonald2011,Carbotte2013,Nicol2014}
\begin{equation}\label{MOC}
\sigma_{ij}(\omega,B)=\frac{i g}{2\pi l_{B}^{2}}\sum_{nm}\frac{f_{n}-f_{m}}{E_{n}-E_{m}}\frac{\langle\psi_{n}\vert j_{i}\vert\psi_{m}\rangle \langle\psi_{m}\vert j_{j}\vert\psi_{n}\rangle}{\hbar\omega + E_{n} - E_{m} + i\eta},
\end{equation}
where $g$ is a degeneracy factor (e.g., $g = 4$ for graphene due to the twofold spin and twofold valley degeneracies),  $l_{B}=\sqrt{\hbar/(eB)}$ is the magnetic length, $\hbar\omega$ is the photon energy, $f_{n}=\{\exp[(E_{n}-\mu)/k_{B}T] + 1\}^{-1}$ is the Fermi-Dirac distribution at temperature $T$ and chemical potential $\mu$, $\eta$ is a small residual scattering rate of charge carriers and $\bm{j}=(ie/\hbar)[\mathcal{H},\bm{r}]$ is the current operator. From Eq. (\ref{AHB}), the $\alpha$-component of $\bm{j}$ results
\begin{equation}
j_{\alpha}=
(ie/\hbar)\sum_{i,j} \tau_{i}v_{ij}[\Pi_{j},r_{\alpha}]=e\sum_{i} v_{\alpha i}\tau_{i},
\end{equation}
for which $[A_{j},r_{\alpha}]=0$, $[p_{j},r_{\alpha}]=-i\hbar\delta_{j\alpha}$ and $v_{i\alpha}=v_{\alpha i}$ were taken into account. Thus, we express the current operator $\bm{j}$ as 
\begin{equation}\label{j}
\bm{j}=e\mbox{\bf{v}}\cdot\bm{\tau}=(\mbox{\bf{v}}\cdot ev_{0}\bm{\tau})/v_{0}=(\mbox{\bf{v}}\cdot \bm{j}^{0})/v_{0},
\end{equation}
where $\bm{j}^{0}=ie[\mathcal{H}^{0},\bm{r}]=ev_{0}\bm{\tau}$ is the current operator of the isotropic 2D DWM taken as reference. Now, in Eq.~(\ref{MOC}) the product of current matrix elements $\langle\psi_{n}\vert j_{i}\vert\psi_{m}\rangle$, and $\langle\psi_{m}\vert j_{j}\vert\psi_{n}\rangle$, which capture the selection rules for Landau-level transitions, can be rewritten using Eq.~(\ref{j}) as
\begin{eqnarray}\label{cme}
\langle\psi_{n}\vert j_{i}\vert\psi_{m}\rangle \langle\psi_{m}\vert j_{j}\vert\psi_{n}\rangle
=
\frac{1}{v_{0}^{2}}\sum_{kl}
\langle\psi_{n}\vert v_{ik}j_{k}^{0}\vert\psi_{m}\rangle
\langle\psi_{m}\vert v_{jl}j_{l}^{0}\vert\psi_{n}\rangle.
\end{eqnarray}
Substituting Eq.~(\ref{cme}) into Eq.~(\ref{MOC}), we obtain
\begin{eqnarray}
\sigma_{ij}(\omega,B)&=&\frac{i g}{2\pi l_{B}^{2}}\frac{1}{v_{0}^{2}}\sum_{nm}\sum_{kl}\frac{f_{n}-f_{m}}{E_{n}-E_{m}}\frac{\langle\psi_{n}\vert v_{ik}j_{k}^{0}\vert\psi_{m}\rangle \langle\psi_{m}\vert v_{jl}j_{l}^{0}\vert\psi_{n}\rangle}{\hbar\omega + E_{n} - E_{m} + i\eta},\nonumber \\
&=&\frac{l_{\mathfrak{B}}^{2}}{l_{B}^{2} v_{0}^{2}}\sum_{kl}v_{ik}\left( \frac{i g}{2\pi l_{\mathfrak{B}}^{2}}\sum_{nm}\frac{f_{n}-f_{m}}{E_{n}-E_{m}}\frac{\langle\psi_{n}\vert j_{k}^{0}\vert\psi_{m}\rangle \langle\psi_{m}\vert j_{l}^{0}\vert\psi_{n}\rangle}{\hbar\omega + E_{n} - E_{m} + i\eta}\right)v_{lj},
\end{eqnarray}
where we conveniently introduce the effective magnetic length $l_{\mathfrak{B}}=\sqrt{\hbar/(e\mathfrak{B})}$. Notice that the term enclosed between parentheses is just the magneto-optical conductivity tensor $\bm{\sigma}^{0}(\omega,\mathfrak{B})$ of an isotropic 2D DWM under the effective magnetic field $\mathfrak{B}\bm{e}_{z}$. Hence, the magneto-optical conductivity tensor of an anisotropic 2D DWM described by Eq.~(\ref{AHB}) is given by the tensorial equation
\begin{equation}\label{main}
\sigma_{ij}(\omega,B)=\frac{1}{\mbox{det}(\mbox{\bf{v}})}\sum_{kl}v_{ik}\sigma_{kl}^{0}(\omega,\mathfrak{B})v_{lj},
\end{equation}
which reads $\bm{\sigma}(\omega,B)=\mbox{\bf{v}}\cdot\bm{\sigma}^{0}(\omega,\mathfrak{B})\cdot\mbox{\bf{v}}/\mbox{det}(\mbox{\bf{v}})$ in a compact notation. The last equation constitutes the main contribution of this paper and allows an efficient determination of $\bm{\sigma}(\omega,B)$ for an anisotropic 2D DWM by means of $\bm{\sigma}^{0}(\omega,\mathfrak{B})$ corresponding to an isotropic one but in a different magnetic field $\mathfrak{B}$ given by Eq.~(\ref{EB}).

\section{Discussion}\label{Dis}

For any reference systems $\bm{\sigma}^{0}(\omega,\mathfrak{B})$ is an antisymmetric tensor of the form \cite{MacDonald2011,Carbotte2013,Nicol2014}
\begin{equation}
\left(\begin{array}{cc}
\sigma_{xx}^{0}(\omega,\mathfrak{B}) & \sigma_{xy}^{0}(\omega,\mathfrak{B})\\
-\sigma_{xy}^{0}(\omega,\mathfrak{B}) & \sigma_{xx}^{0}(\omega,\mathfrak{B})
\end{array}\right),
\end{equation}
which has only two independent components $\sigma_{xx}^{0}(\omega,\mathfrak{B})$ and $\sigma_{xy}^{0}(\omega,\mathfrak{B})$, instead of at least three ones for an anisotropic 2D DWM to be calculated using the Kubo formula (\ref{MOC}). These two components respectively denote the longitudinal and Hall conductivities of an isotropic material. Then, from Eq.~(\ref{main}) the four components of $\bm{\sigma}(\omega,B)$ can be explicitly written as
\begin{eqnarray}
\sigma_{xx}(\omega,B)&=& \left(\frac{\mbox{tr}(\mbox{\bf{v}})}{\mbox{det}(\mbox{\bf{v}})}v_{xx}-1\right)\sigma_{xx}^{0}(\omega,\mathfrak{B}),\label{XX}\\
\sigma_{yy}(\omega,B)&=& \left(\frac{\mbox{tr}(\mbox{\bf{v}})}{\mbox{det}(\mbox{\bf{v}})}v_{yy}-1\right)\sigma_{xx}^{0}(\omega,\mathfrak{B}),\label{YY}\\
\sigma_{xy}(\omega,B)&=& \sigma_{xy}^{0}(\omega,\mathfrak{B}) + \frac{\mbox{tr}(\mbox{\bf{v}})}{\mbox{det}(\mbox{\bf{v}})}v_{xy}\sigma_{xx}^{0}(\omega,\mathfrak{B}),\label{XY}\\
\sigma_{yx}(\omega,B)&=& -\sigma_{xy}^{0}(\omega,\mathfrak{B}) + \frac{\mbox{tr}(\mbox{\bf{v}})}{\mbox{det}(\mbox{\bf{v}})}v_{xy}\sigma_{xx}^{0}(\omega,\mathfrak{B}).\label{YX}
\end{eqnarray}

In other words, the magneto-optical conductivity tensor of an anisotropic 2D DWM, generally characterized by four components, is fully determined by calculating only $\sigma_{xx}^{0}(\omega,\mathfrak{B})$ and $\sigma_{xy}^{0}(\omega,\mathfrak{B})$ of an arbitrary isotropic 2D DWM immersed in an effective magnetic field $\mathfrak{B}$ given by Eq.~(\ref{EB}). 

From Eqs. (\ref{XY}) and (\ref{YX}), it is clear that in an arbitrary laboratory reference system $\bm{\sigma}(\omega,B)$ is a tensor without defined symmetry, i.e., it is neither symmetric nor antisymmetric. However, in the principal coordinate system $x'y'$, where $\mbox{\bf{v}}$ is diagonal with $v_{x'y'}=0$, $\bm{\sigma}(\omega,B)$ results antisymmetric, whose components are given by the simplified expressions  
\begin{eqnarray}
\sigma_{x'x'}(\omega,B) &=& \frac{v_{x'x'}}{v_{y'y'}}\sigma_{xx}^{0}(\omega,\mathfrak{B}),\\
\sigma_{y'y'}(\omega,B) &=& \frac{v_{y'y'}}{v_{x'x'}}\sigma_{xx}^{0}(\omega,\mathfrak{B}),\\
\sigma_{x'y'}(\omega,B) &=& -\sigma_{y'x'}(\omega,B)=\sigma_{xy}^{0}(\omega,\mathfrak{B}),
\end{eqnarray} 
after evaluating Eqs. (\ref{XX}--\ref{YX}) for the reference system $x'y'$. 

For the limiting case $B=0$, one has $\sigma_{xy}^{0}(\omega,0)=0$. Thus, the conductivity tensor $\bm{\sigma}^{0}(\omega,0)$ takes the simple form $\bm{\sigma}^{0}(\omega,0)=\sigma^{0}(\omega)\mbox{\bf{I}}$, where $\mbox{\bf{I}}$ is the $(2\times2)$ identity matrix and $\sigma^{0}(\omega)\equiv\sigma^{0}_{xx}(\omega,0)$ is the optical conductivity of an isotropic 2D DWM in the absence of magnetic field. Consequently, for $B=0$ the tensorial expressions (\ref{XX}--\ref{YX}) become 
\begin{equation}\label{CsB}
\bm{\sigma}(\omega)=\sigma^{0}(\omega)\left(\frac{\mbox{tr}(\mbox{\bf{v}})}{\mbox{det}(\mbox{\bf{v}})}\mbox{\bf{v}} - \mbox{\bf{I}}\right),
\end{equation}
which captures the optical conductivity tensor of an anisotropic 2D DWM described by (\ref{AH}) in the absence of magnetic field.

\section{Application to strained graphene}\label{ASG}

As an example to illustrate the obtained results, let us consider an uniformly strained graphene. Up to first order in the strain tensor $\bm{\epsilon}$, the dynamics of its low-energy carriers can be described by Dirac-Weyl Hamiltonian (\ref{AH}) with a Fermi velocity tensor given by \cite{Pellegrino2011,Oliva2013,Oliva2017} 
\begin{equation}\label{vsg}
\mbox{\bf{v}}=v_{F}(\mbox{\bf{I}}-\beta\bm{\epsilon}),
\end{equation}
where $\beta\sim2$ and $v_{F}$ is the Fermi velocity for unstrained graphene, being the chosen isotropic reference material. Substituting Eq.~(\ref{vsg}) into Eqs.~(\ref{XX}--\ref{YX}), replacing $v_{0}$ by $v_{F}$ and linearizing with respect to $\bm{\epsilon}$, the magneto-optical conductivity of strained graphene under a magnetic field $B\bm{e}_{z}$ is given by
\begin{eqnarray}
\sigma_{xx}(\omega,B)&=& \big[1-\beta(\epsilon_{xx}-\epsilon_{yy})\big]\sigma_{xx}^{0}(\omega,\mathfrak{B}),\label{XXsg}\\
\sigma_{yy}(\omega,B)&=& \big[1-\beta(\epsilon_{yy}-\epsilon_{xx})\big]\sigma_{xx}^{0}(\omega,\mathfrak{B}),\label{YYsg}\\
\sigma_{xy}(\omega,B)&=& \sigma_{xy}^{0}(\omega,\mathfrak{B}) - 2\beta\epsilon_{xy}\sigma_{xx}^{0}(\omega,\mathfrak{B}),\label{XYsg}\\
\sigma_{yx}(\omega,B)&=& -\sigma_{xy}^{0}(\omega,\mathfrak{B}) - 2\beta\epsilon_{xy}\sigma_{xx}^{0}(\omega,\mathfrak{B}),\label{YXsg}
\end{eqnarray}
where $\mathfrak{B}=B\big[1-\beta\mbox{tr}(\bm{\epsilon})\big]$ is derived from Eq.~(\ref{EB}) and $\bm{\sigma}^{0}(\omega,\mathfrak{B})$ denotes the magneto-optical conductivity of unstrained graphene, which has been theoretically \cite{Gusynin2007,Ferreira2011} and experimentally \cite{Sadowski2006,Jiang2007} addressed.

\begin{figure}[t]
\centering
\includegraphics[width=0.7\linewidth]{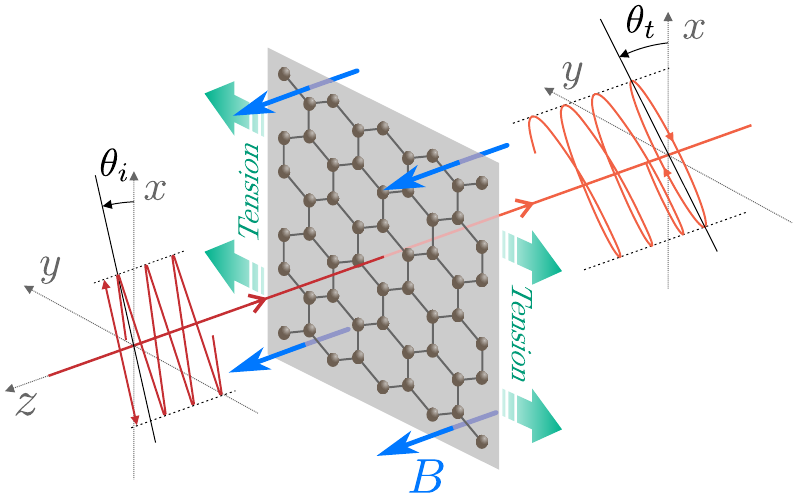}
\caption{\label{fig2} Schematic representation of a generalized Faraday polarization rotation experiment: a linearly polarized light beam becomes elliptically polarized after passing strained graphene.}
\end{figure}

In order to illustrate the effects of both anisotropy and magnetic field on the optical properties, let us analyse the transmittance of linearly polarized light for normal incidence on strained graphene, as shown in Fig.~\ref{fig2}. Considering strained graphene as a two-dimensional sheet with conductivity $\bm{\sigma}(\omega,B)$ given by Eqs.~(\ref{XXsg}--\ref{YXsg}) and from the boundary conditions for the electromagnetic field along the interface vacuum-graphene-vacuum, one can demonstrate that the electric fields of incident and transmitted waves, $\bm{E}_{i}$ and $\bm{E_{t}}$, are related by \cite{Stauber2008,Oliva2015}
\begin{equation}\label{EiEt}
\bm{E}_{i}=\left(\mbox{\bf{I}}+\frac{1}{2}Z_{0}\bm{\sigma}(\omega,B) \right)\cdot\bm{E}_{t},
\end{equation}
where $Z_{0}$ is the impedance of vacuum. Notice that for $\mbox{Im} [\bm{\sigma}(\omega,B)]\neq0$ the transmitted light beam acquires a certain ellipticity. 

\begin{figure}[ht]
\centering
\includegraphics[width=0.56\linewidth]{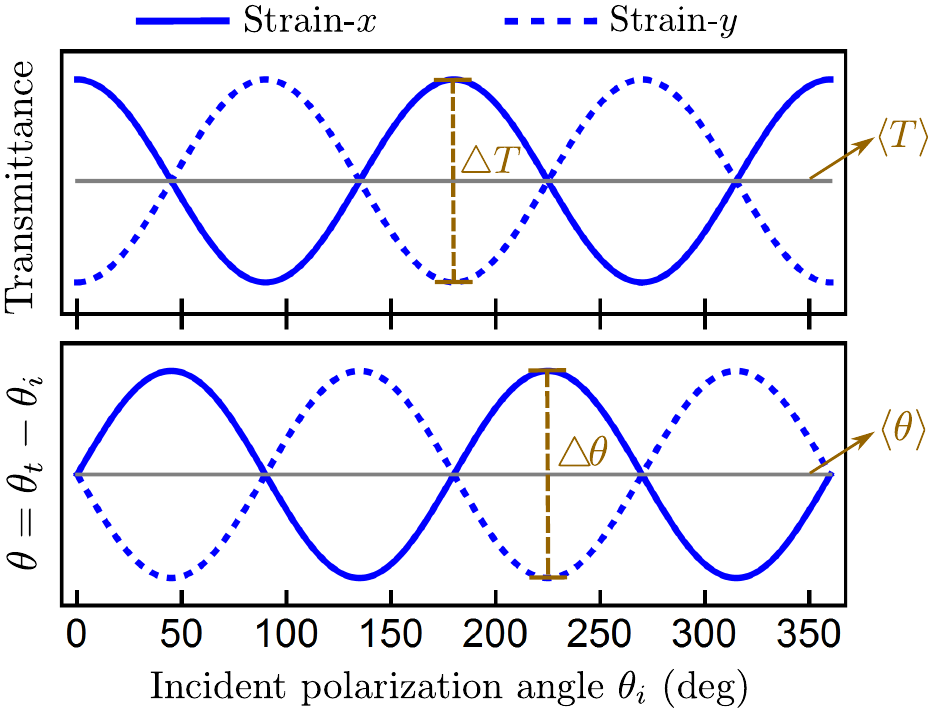}
\caption{\label{fig3} Transmittance (top panel) and polarization rotation (bottom panel) as a function of the incident polarization angle $\theta_{i}$ for a uniaxial strain along the $x$-axis ($y$-axis) of the laboratory reference system, illustrated by solid (dashed) curves. Both strains have the same magnitude.}
\end{figure}

In the limit of weak absorption, from Eq.~(\ref{EiEt}) it is straightforward to write the transmittance $T$ as \cite{Oliva2015}
\begin{equation}\label{Tsg}
T(\omega,B,\theta_{i})\approx 1-Z_{0}\mbox{Re}[\sigma_{xx}^{0}(\omega,\mathfrak{B})] 
\Big(1+\beta(\epsilon_{yy}-\epsilon_{xx})\cos2\theta_{i}-2\beta\epsilon_{xy}\sin2\theta_{i}\Big),
\end{equation}
where $\theta_{i}$ denotes the incident polarization angle, as illustrated in Fig.~\ref{fig2}. This expression reveals a periodic modulation of the transmittance as a function of $\theta_{i}$ due to the anisotropy of strained graphene (see Fig.~\ref{fig3}). Such modulation has been observed by Ni \emph{et al.} \cite{Ni2014}, but in absence of magnetic field. Now an external magnetic field, through the longitudinal conductivity $\sigma_{xx}^{0}(\omega,\mathfrak{B})$,  can modify both the mean value of the transmittance $\langle T\rangle =  1-Z_{0}\mbox{Re}[\sigma_{xx}^{0}(\omega,\mathfrak{B})]$ and the amplitude of the modulation $\triangle T=2\beta Z_{0}\mbox{Re}[\sigma_{xx}^{0}(\omega,\mathfrak{B})][\mbox{tr}(\bm{\epsilon})^{2}-4\mbox{det}(\bm{\epsilon})]^{1/2}$. 

Unlike the transmittance, the polarization rotation of a linearly polarized light beam after passing strained graphene depends on both the longitudinal conductivity $\sigma_{xx}^{0}(\omega,\mathfrak{B})$ and the Hall conductivity $\sigma_{xy}^{0}(\omega,\mathfrak{B})$.  From Eq.~(\ref{EiEt}) and in the limit of weak absorption, the polarization rotation angle $\theta=\theta_{t}-\theta_{i}$ results
\begin{equation}\label{FA}
\theta(\omega,B,\theta_{i})\approx\frac{1}{2}Z_{0}\mbox{Re}[\sigma_{xy}^{0}(\omega,\mathfrak{B})] +\beta Z_{0}\mbox{Re}[\sigma_{xx}^{0}(\omega,\mathfrak{B})]
\left(\epsilon_{xy}\cos2\theta_{i} + \frac{\epsilon_{yy}-\epsilon_{xx}}{2}\sin2\theta_{i}\right),
\end{equation}
where $\theta_{t}$ is the transmitted polarization angle (see Fig.~\ref{fig2}).

In Eq.~(\ref{FA}) one can separately identify the effects of either magnetic field or strain-induced anisotropy. The first term in its right side, $Z_{0}\mbox{Re}[\sigma_{xy}^{0}(\omega,\mathfrak{B})]/2$, is owing to the Faraday effect. In fact, this term has the same form of the Faraday rotation angle for unstrained graphene \cite{Crassee2010,Fialkovsky2012}. On the other hand, the second term in the right side of Eq.~(\ref{FA}), which is dependent on $\theta_{i}$, is essentially due to the strain-induced anisotropy. Even in absence of the magnetic field, this second term survives and describes the dichroism of strained graphene reported in Ref.~\cite{Oliva2015}. The main difference between these two terms of Eq.~(\ref{FA}) is the dependence on the incident polarization direction. While the first term related to the Faraday effect does not dependent on the incident polarization angle $\theta_{i}$, the strain-induced term is $\theta_{i}$-dependent. For instance, if the incident polarization direction is collinear to one of the principal directions of the strain tensor, the second term is equal to zero.

In summary, the polarization rotation angle given by Eq.~(\ref{FA}) can be recognized as the generalized expression of the Faraday rotation angle for graphene under uniform strain. Its remarkable feature respect to the version for unstrained graphene is the periodic variation as a function of the incident polarization direction. The mean value of the generalized Faraday rotation angle is given by $\langle\theta\rangle\approx Z_{0}\mbox{Re}[\sigma_{xy}^{0}(\omega,\mathfrak{B})]/2$, whereas the amplitude of its modulation is $\triangle \theta=\beta Z_{0}\mbox{Re}[\sigma_{xx}^{0}(\omega,\mathfrak{B})][\mbox{tr}(\bm{\epsilon})^{2}-4\mbox{det}(\bm{\epsilon})]^{1/2}$, which are represented in Fig.~\ref{fig3}.

\section{Generalization for arbitrary pseudospin}\label{Gen}

\begin{figure}[ht]
\centering
\includegraphics[width=0.9\linewidth]{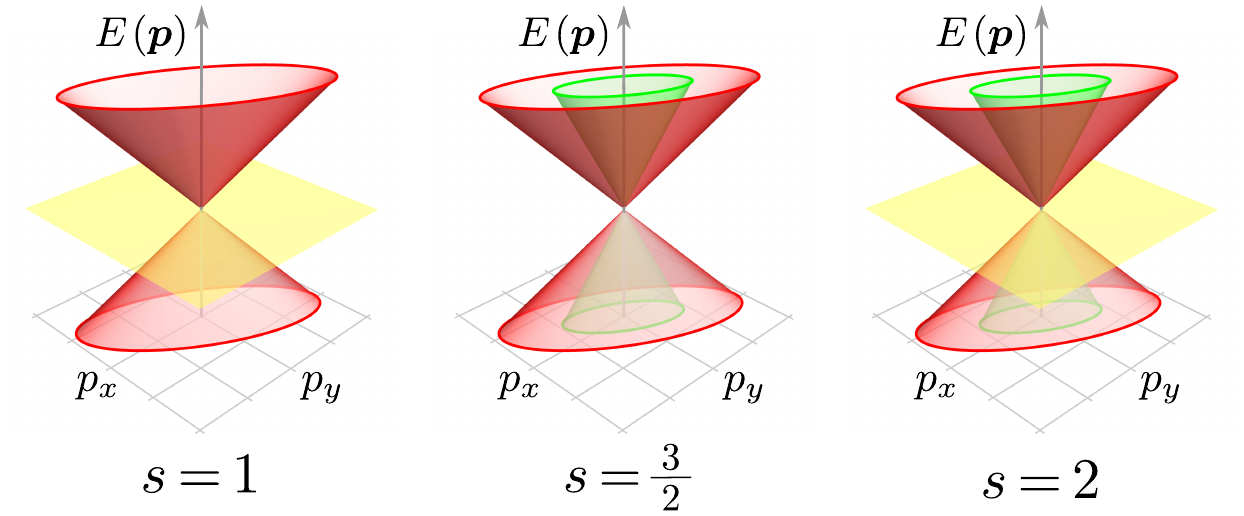}
\caption{\label{fig4} Energy dispersion relations for anisotropic 2D Dirac-Weyl materials with pseudospin-$s$ equal to $1$, $3/2$, and $2$. The yellow planes represent zero-energy flat bands.}
\end{figure}

Up to this point, we have studied the magneto-optical conductivity of anisotropic 2D DWMs for the case of pseudospin, $s=1/2$, as occurred in graphene and topological insulator. Reviewing in detail the derivation of Eq.~(\ref{main}), two requirements have been used: (i) The current operators of isotropic and anisotropic 2D DWMs are related by Eq.~(\ref{j}). (ii) The anisotropic 2D DWM in an external magnetic field $B$ has exactly the same Landau level spectrum of the isotropic one under an effective magnetic field $\mathfrak{B}$ given by Eq.~(\ref{EB}). These two requirements are also hold in the general case of an anisotropic 2D DWM described by a Hamiltonian of the form
\begin{equation}\label{GADWH}
\mathcal{H}=\bm{s}\cdot\mbox{\bf{v}}\cdot\bm{p},
\end{equation}
where $\bm{s}=(s_{x},s_{y})$ are the first two spin-$s$ matrices, being $s$ integer or half-integer. The corresponding energy dispersion relation consists $2s+1$ bands, in the form of nested deformed Dirac cones and zero-energy flat band if $s$ is an integer, as illustrated in Fig.~\ref{fig4}.

Thus, proceeding analogously as we have made for the case $s=1/2$, one can demonstrate that Eq.~(\ref{main}) is also valid for anisotropic 2D DWMs with carriers of arbitrary pseudospin-$s$. 

At zero magnetic field, the optical properties of isotropic 2D DWMs with arbitrary pseudospin-$s$ have been previously calculated \cite{Dora2011}. Now, using Eq.~(\ref{vsg}) the findings obtained by D\'ora \emph{et al.} \cite{Dora2011} can be extended for 
the case of anisotropic DWM. More recently, the magneto-optical conductivity of isotropic systems, that obey the general pseudospin-$s$ 2D Dirac-Weyl Hamiltonian, has been evaluated in Ref.~\cite{Malcolm2014}, where the authors particularly focused on $s = \{1/2,1,3/2,2\}$ and showed that the magneto-optical behaviours are markedly different for each case considered. Then, our Eq.~(\ref{main}) permits a generalization of the results reported by Malcolm \emph{et al.} \cite{Malcolm2014} to address the magneto-optical response of \emph{anisotropic} 2D DWM with arbitrary pseudospin-$s$.

\section{Conclusions}\label{Con}

Starting from Kubo formalism, we derived an analytical expression for the magneto-optical conductivity tensor of a generic anisotropic (strained) 2D DWM. For an isotropic 2D DWM under an external magnetic field, such tensor is symmetric, in contrast to an antisymmetric one for an anisotropic 2D DWM in absence of magnetic field. However, when both the magnetic field and anisotropy are present, as considered in this work, the optical conductivity of a 2D DWM is given by a neither symmetric nor antisymmetric tensor, which is unusual in solid state physics.

Tensorial equation (\ref{main}) essentially enables an easy access to the complex magneto-optical conductivity of anisotropic 2D DWMs from the well known magneto-optical response of isotropic ones. As an example, we have applied our results to the case of strained graphene. In particular, we quantified by means of analytical expressions the effects of strain-induced anisotropy and magnetic field on the light transmittance and on the polarization rotation. Moreover, from the generalized expression (\ref{FA}) of the Faraday rotation angle, we were able to identify the strain-induced effects as in comparison to the magnetic effects. The former is dependent on the incident polarization direction. Finally, we discussed the magneto-optical conductivity tensor for anisotropic 2D DWMs with arbitrary pseudospin.

\begin{acknowledgments}
This work has been partially supported by CONACyT of Mexico through Project 252943, and by PAPIIT of Universidad Nacional Aut\'onoma de M\'exico (UNAM) through Project IN106317. M.O.L. acknowledges a postdoctoral fellowship from DGAPA-UNAM.
\end{acknowledgments}

\bibliography{biblioStrainedGraphene}

\end{document}